# Effect of convective transport in edge/SOL plasmas of ADITYA-U tokamak

Ritu Dey[1], Joydeep Ghosh[1,2], Tanmay M. Macwan[1,2,*], Kaushlender Singh[1,2], M. B. Chowdhuri[1], H. Raj[1], R. L. Tanna[1], Deepti Sharma[1], T. D. Rognlien[3]

[1] Institute for Plasma Research, Bhat, Gandhinagar 382 428, India
[2] Homi Bhaba National Institute, Training School Complex, Anushakti Nagar, Mumbai 400 094, India
[3] Lawrence Livermore National Laboratory, Livermore, CA 9455, USA

E-mail: ritu.dey@ipr.res.in
*Currently working at University of California, Los Angeles, CA 90095, USA

**Abstract**

The 2-D edge plasma fluid transport code, UEDGE has been used to simulate the edge region of circular limiter plasmas of ADITYA-U for modelling the measured electron density profile. The limiter geometry of ADITYA-U has been introduced in the UEDGE code, which is primarily developed and used for divertor configuration. The computational mesh defining the limiter geometry is generated by a routine developed in-house, and has successfully been integrated with the UEDGE code to simulate the edge plasma parameters of ADITYA-U. The radial profiles of edge and scrape-off layer (SOL) electron density, $n_e$ and temperature are obtained from the simulations and used to model the measured $n_e$ profile using Langmuir probe array. It has been found that a convective velocity, $v_{conv.}$ is definitely needed in addition to the constant perpendicular diffusion coefficient, D throughout the edge and SOL regions to model the edge $n_e$ profile. The obtained $v_{conv.}$ is inward and radially constant with a value of 1.5 m/s and the radially constant D is ~ 0.2 m$^2$/s. The value of D ~ 0.2 m$^2$/s is found to be much less than fluctuation induced diffusivities and lies in-between the neoclassical diffusivity and Bohm diffusivity estimated in the edge-SOL region of ADITYA-U tokamak. Furthermore, the transport of radial electron heat flux is found to be maximizing near the limiter tip location in the poloidal plane.

## 1. Introduction

The edge plasma region of a tokamak, i.e., the plasma region on either side of the last closed flux surface (LCFS) can be divided into two sub-regions, i) the core-edge region located inside the LCFS and ii) the scrape-off-layer (SOL) region located outside the LCFS [1]. In the core-edge region, the magnetic field lines are closed on themselves, and charged particles move along the field lines and are confined to the flux surfaces. Whereas, outside of the LCFS, the open magnetic field lines remain in direct contact with material walls/limiter/divertor of the fusion reactor. In this region, the particles leave the plasma by coming in contact with the wall/limiter/divertor due to fast motion along field lines within a short time [2, 3]. As a result, the charged-particle confinement remains poor as compared to the core-edge region. Understanding the edge region of the plasma of any tokamak is very important as it isolates the hot core plasma from the metal boundary of the machine. The properties of this region determine the transport of particles and energy in and out of the core plasma and decide the fate of the both core-plasma as well as of the material boundary in terms of their operating lifetime. The dynamics of this region are very complicated due to the presence of sharp gradients of density, temperature, and radial electric fields, enhanced level of fluctuations, open-magnetic field lines terminating on the material surfaces, etc. [2]. The presence of various time-scales and length-scales associated with the transport of ions and electrons along and across the confining/open magnetic fields, together with the neutral particle transport further complicate the modelling of this region [4, 5]. Furthermore, various atomic and molecular processes of different fuel and impurity species take place in this region as a result of the interaction of plasma with a neutral gas, impurities, and material surfaces such as divertor targets/limiters [6, 7].

Exhaustive measurements of various quantities in the edge region of any tokamak with good spatial and temporal resolution have their own challenges and limitations. Hence, simulation studies are essential to adequately model this region and to understand the underlying physical processes in-toto. Several simulations studies are employed in different tokamaks to predict the plasma parameters with the help of various edge plasma transport codes e.g., SOLPS [8-12], EDGE2D-EIRENE [13, 14], and UEDGE [15]. SOLPS is applied rigorously to interpret the experimental measurements of ASDEX-Upgrade [8], HT-7 [9], EAST [10], TCV [11], and JET [12] tokamaks. The JET plasma is modelled using EDGE2D-EIRENE code and the experimental profile of the plasma parameters are compared with the simulation results and in good agreement [13].

The plasma edge transport modelling requires mostly the application of the anomalous diffusion coefficient. Its values are either constant or radially varying depending upon devices and plasma properties. In Alcator C-Mod tokamak, M. V. Umansky *et al* [16] have studied the recycling and transport in the edge plasma using the UEDGE code and found that spatially varying D ~ 1.0 to 0.1 $m^2$/s is necessary to model the electron density in SOL region at mid-plane gas pressure 0.025 mTorr. Matching the experimental plasma density profiles in the scrape-off layer (SOL) requires a spatially dependent effective anomalous diffusion coefficient D growing rapidly towards the wall. Recently, the radially varying diffusion coefficient is also used to explain the experimental measurements [17] in ITER with SOLPS-ITER code. SOLPS based numerical simulation has been done for the evolution of the energy deposition on the divertor target of EAST [18] and perpendicular anomalous transport coefficients (D) is adjusted to match the upstream experimental plasma parameters. Here, they have considered the radially varying D ⁽⁾ ⁽⁾Wensing et al [19] is performed a quantitative test of SOLPS-ITER simulations of L-mode experiments in TCV, where they have used spatially constant anomalous transport coefficient. In the ASDEX-Upgrade H-mode plasma, a radially varying diffusion coefficient is taken into consideration to reproduce the experimentally measured electron density profile using SOLPS code [8].

It is to be noted that the diffusive transport is only taken into account to simulate the experimental measurements in the above described studies. However, although not in many tokamaks, non-diffusive transport is also sometimes required to understand the edge-SOL physics in the tokamaks. As for example, fast anomalous cross-field plasma transport in the SOL region of DIII-D tokamak is observed through the modelling using the UEDGE code [20]. In that study they have used a spatially dependent, time-averaged anomalous convective velocity and found that inclusion of convective velocity not only reproduced the individual properties of the selected shots but also some important experimental features like the extension of radial density profile far into SOL region. Furthermore, depending on various plasma conditions in DIII-D [21], it has been observed that the simulated edge plasma parameters are well matched with the experimental measurements in SOL and the divertor region after including maximum convective velocity in the range of 50 to 100 m/s along with the diffusion coefficient in the range of 0.125-0.3 $m^2/s$. Barbora Gulejova [22] presented the simulated upstream profiles with SOLPS code modelling by including radially varying D, χ and convective velocity. The values of convective velocity that have been found to be required to achieve agreement between code and experiment are interestingly rising functions of radial distance in the SOL (about from 0 to 30 m $s^{-1}$). It is mentioned that, while modelling the experimental observations from TCV-SOL, the convective velocity is needed to be included to describe the turbulent perpendicular transport, especially in high-density region.

The above studies primarily concern the divertor configuration and not for limiter configuration. Although, the tokamak plasmas with limiter configuration are studied for many years but it is yet to be fully established as one can see that the obtained transport parameters through the modelling of edge plasma parameter are varied from devices to devices and depending upon their operational regimes. In addition to that, the studies of the tokamak edge plasma with limiter configuration becomes more relevant in the viewpoint of fusion-grade ITER tokamak since its plasma operation is planned to be performed without the blanket modules and with temporary stainless steel limiters [23]. In Ref [24] it is mentioned that during the subsequent engineering operation phase for commissioning of the magnet systems to full current, circular limiter plasmas of up to 1 MA might be attempted. The experimental temperature and density profiles in HT-7 with limiter configuration is modelled using SOLPS code by varying the perpendicular particle transport coefficient D and heat conductivities $x_e$, in the code until good agreement between the code results and the measured data has been achieved. They have used step function like diffusion coefficient where D varies from ~1.4 to 1.1 $m^2$/sec. Along with the above, using UEDGE code, the edge plasma modelling is done for ITER with limiter configuration along with the divertor and the limiter is penetrated upto seperatrix [25]. It is found that the distribution of total heat load between the divertor plates and outer mid-plane surfaces can be controlled by the limiter penetration depth [25]. In this work, the diffusion coefficient is taken as spatially constant and the value is 0.33 m$^2$/s.

Presently, the UEDGE code simulations are being used to comprehend the edge and SOL plasma behaviours in the discharges of the ADITYA-U tokamak with limiter configuration. As the UEDGE code is primarily developed for divertor geometry, the code is adequately modified for implementing it in a limiter geometry. The computational-mesh incorporating the limiter geometry for ADITYA-U is generated using a MATLAB/PYTHON based mesh-generator based on plasma equilibrium flux surfaces [26] and integrated successfully into the UEDGE code. In earlier studies, the computational-mesh for poloidal ring-limiter geometry of ADITYA tokamak is successfully integrated into the neutral particle transport code (DEGAS2) and neutral particle dynamics have been thoroughly investigated [27, 28, 29].

After setting up the UDEGE code for the limiter geometry of ADITYA-U, the radial profiles of electron density and temperature in the edge-SOL region are obtained. These simulated values are then compared with the Langmuir probe measurements to understand the transport properties of particles and energy in the edge-SOL region. It has been observed that the simulated and measured radial electron density profiles in the edge-SOL region could not be matched well by applying only constant particle diffusion coefficient (D) of ~ 0.2 m$^2$/s in the whole edge-SOL region. For achieving the best-matched condition, it is necessary to apply a constant diffusion coefficient together with an inward convection velocity that is radially constant. Interestingly, the obtained value of particle diffusion coefficient through comparison of simulated and measured density profiles is found to be higher than the estimated neoclassical value. However, the value of D is found to be smaller than the estimated Bohm diffusion [30] and significantly smaller than the fluctuation induced diffusivity [31, 32]. Furthermore, it has been observed that the convective and conductive processes contribute to electron heat fluxes having maximum values of ~ 1.3 × 10$^{21}$ eV m$^{-2}$ s$^{-1}$ and ~ 2.1 × 10$^{21}$ eV m$^{-2}$ s$^{-1}$, respectively. The electron conductive heat-flux remains higher compared to the convective heat flux till a distance of ~ - 0.5 cm from the limiter, however, the total radial electron heat flux becomes maximum near the limiter tip location.

The paper is organized as follows: In Section 2, the ADITYA tokamak's machine and discharge parameters as well as the relevant diagnostics are presented. Section 3 describes the modified UEDGE code for simulating the edge plasma region of ADITYA-U tokamak plasmas in the limiter configuration. The simulation results and comparison with experimental measurements are presented in section 4. The paper is concluded in section 5.

## 2. ADITYA tokamak and relevant diagnostics

The ADITYA-U [33,34], a medium-sized tokamak, is having a toroidal belt limiter at the inner (high toroidal magnetic field) side and two poloidal limiters (toroidal widths of ~ 8 cm) at the outer (low toroidal magnetic field) side located 180° apart [35]. The plasma major and minor radii are 0.75 m and 0.25 m, respectively. The typical discharges of ADITYA-U have plasma current ($I_p$) ~ 100 - 200 kA, central chord-averaged density $n_e$ ~ 1 - 4× $10^{19}$ m$^{-3}$, $B_T$ ~ 1.0 - 1.44 T, and chord-averaged central electron temperature ~ 250 - 500 eV. The base pressure is ~ 6 - 9 × $10^{-9}$ Torr, and the hydrogen discharges are produced at a pre-fill gas pressure of ~ 1 - 2 × $10^{-4}$ Torr [33]. The edge plasma region of ADITYA-U is equipped with different types of Langmuir probes that measure the plasma parameters in this region [36]. These are placed at different radial and poloidal locations in the edge-SOL region of ADITYA-U. The probes are composed of molybdenum, and their exposed surfaces have a half-spherical shape with a diameter. Since each probe is housed inside a non-machinable ceramic (alumina, $Al_2O_3$) with a distinct radial length, they are all distributed along the edge/SOL at various radial places [37]. The central chord averaged electron density is measured by single channel heterodyne $\tilde{m}$wave interferometer diagnostics. The radial profile of electron density is measured by seven channels homodyne $\tilde{m}$wave interferometer spanning over whole diameter of the plasma. The core plasma electron temperature is estimated from the soft X-ray emissions using the foil absorption technique.

## 3. Implementation of UEDGE code for limiter discharges of ADITYA-U tokamak

### 3.1 Theory

The UEDGE code [15] is an edge-plasma fluid transport code for determining the plasma parameters such as plasma density, ion velocity (parallel to the magnetic field), ion and electron temperatures, and electrostatic potential of the edge-SOL plasmas of tokamaks. The fuel as well as impurity neutrals, in the UEDGE, are introduced using two models, one is an inertial fluid model and another is the diffusive model. Additionally, the neutrals can be introduced using a Monte Carlo neutral model, used in DEGAS2 neutral code. The basic physics equations are taken from Braginskii [38], with the addition of anomalous or turbulence-enhanced transport coefficients for the direction across the magnetic field whereas the transport along the magnetic field line is taken as classical with flux limits. The dynamical fluid equations for particle continuity, parallel (to the B-field) ion velocity, and separate ion and electron temperatures which are considered in the UEDGE code are given below.

The continuity equation for ions can be written as:

$$\frac{\partial}{\partial t}n_i + \frac{1}{V}\frac{\partial}{\partial x}\left(\frac{V}{h_x}n_i u_{ix}\right) + \frac{1}{V}\frac{\partial}{\partial y}\left(\frac{V}{h_y}n_i u_{iy}\right) = <\sigma_i v_{te}> n_e n_g - <\sigma_r v_{te}> n_e n_i,$$

----------(1)

The y-coordinate varies perpendicular to flux surfaces (radial direction) and the x-coordinate varies along the flux surfaces (poloidal direction). The terms $<\sigma_i v_{te}>$ and $<\sigma_r v_{te}>$ are the rate coefficients for ionization and recombination, respectively. The metric coefficients are $h_x = 1/\|\nabla x\|$, $h_y = 1/\|\nabla y\|$, and $V = 2\pi R h_x h_y$ is the volume element for toroidal geometry with major radius R. Rest of the manuscript, the metric coefficients are suppressed in the remaining equations. The rates for the above atomic processes are interpolated from tables compiled as a function of electron temperature and density in the ADAS database [39]. In Eq (1) $n_i$ and $u_x$ are the density and poloidal velocity of ion. The poloidal velocity of the ion is expressed as,

$$u_{ix} = \frac{B_x}{B}v_{i\|} + v_{x,E} + v_{ix,\nabla B}$$

---------(2)

The second term on the R.H.S of equation (2) represents the $(E \times B)/B^2$ drift while the third term represents the drift term due to the sum of curvature and grad B drifts scaled as $qT_i/RB$ (q=ion charge, R=major radius) [15]. The total velocity is denoted by u, whereas the classical component is denoted by v. The parallel velocity is considered classical, so $u_\| = v_\|$. The radial ion velocity is as follows:

$$u_{iy} = -\frac{D_a}{n_i}\frac{\partial n_i}{\partial y} + V_a + v_{y,E} + v_{iy,\nabla B} + v_{iy,vis}$$

---------(3)

where $D_a$ and $V_a$ are the anomalous ambipolar transport coefficients of plasma particles and which characterized turbulence-driven transport. The third term ($v_{y,E}$) is the EXB-drift in the y-direction while $v_{iy,\nabla B}$, is the sum of the curvature and grad-B drifts in y-direction. The last term is an anomalous viscous drift that gives a connection between the electrostatic potential on neighbouring magnetic flux surfaces [15]. The electron velocities have the same expression as the ion velocities except for two differences. The first one is that in the scaling of the $\nabla B$ drift which is $-eT_e/RB$, and the second one is that the electron perpendicular viscosity is neglected due to the much smaller gyro-radii of electrons.

The ion parallel momentum equation is

$$\frac{\partial}{\partial t}(m_i n_i v_{i\|}) + \frac{\partial}{\partial x}\left(m_i n_i v_{i\|} u_{ix} - \eta_{ix}\frac{\partial v_{i\|}}{\partial x}\right) + \frac{\partial}{\partial y}\left(m_i n_i v_{i\|} u_{iy} - \eta_{iy}\frac{\partial v_{i\|}}{\partial y}\right)$$
$$= \frac{B_x}{B}\left(-\frac{\partial P_p}{\partial x}\right) - m_i n_g \nu_{cx}(v_{i\|} - v_{g\|}) + m_g n_g \nu_i v_{g\|} - m_i n_i \nu_r v_{i\|}$$

--------(4)

where, $P_p=P_e+P_i$. $P_p$, $P_e$ and $P_i$ are the total plasma, electron and ion pressure, respectively and $P_{i,e}=n_{i,e} T_{i,e}$.

$\eta_{ix} = \left(\frac{B_x}{B}\right)^2 \eta_\parallel$ is the classical viscosity and $\eta_{iy}$ is the anomalous viscosity. All classical viscosities and thermal velocities are flux-limited [15]. $n_g$ and $v_{g\parallel}$ are the density and parallel velocity of neutral hydrogen, respectively. The last three terms of the equation (4) contain the momentum exchange between neutral and ions via charge-exchange process, neutral and electron via ionization process and between ion and electron via recombination process, respectively. It is to be noted that $m_i=m_g$ for atomic hydrogen neutrals.

The electron energy equation is

$$\frac{\partial}{\partial t}\left(\frac{3}{2}n_e T_e\right) + \frac{\partial}{\partial x}\left[C_{ex}n_e u_{ex}T_e - \kappa_{ex}\frac{\partial T_e}{\partial x} - 0.71 n_e T_e \frac{B_x}{B}\frac{J_\parallel}{en_e}\right] + \frac{\partial}{\partial y}\left[C_{ey}n_e u_{ey}T_e - \kappa_{ey}\frac{\partial T_e}{\partial y}\right]$$
$$= \left[u_{ix}\frac{\partial P_e}{\partial x} - u_{iy}\frac{\partial P_i}{\partial y} - u_{iw}\frac{B_x}{B}\frac{\partial P_p}{\partial x}\right] + \mathbf{E}\cdot\mathbf{J} - K_q(T_e - T_i) + S_{E_e}$$

-------(5)

The first and second square brackets on the L.H.S of equation (5) are the heat fluxes $q_{x,e}$ and $q_{y,e}$, respectively. The first term and second terms of the heat fluxes are the convected and conducted heat fluxes, respectively, in the poloidal and radial directions. Here the poloidal heat conductivity ($k_{ex}$) is classical and the radial heat conductivity is anomalous $k_{ey}=n_e\chi_e$. $C_{ex}$ and $C_{ey}$ are the convection coefficients. The velocity $u_{iw}$ is only used when cross-field drift is considered [15]. $S_{Ee}$ represents the volumetric sinks due to radiation which is the electron energy sink. The ion energy equation which is an implied sum over the ion and neutral species using j indices with $T_g=T_i$ is as follows:

$$\frac{\partial}{\partial t}\left(\frac{3}{2}n_i T_i\right) + \frac{\partial}{\partial x}\left[C_{ix} n_i u_{ix} T_i - \kappa_{jx}\frac{\partial T_i}{\partial x}\right] + \frac{\partial}{\partial y}\left[C_{iy} n_i u_{iy} T_i - \kappa_{jy}\frac{\partial T_i}{\partial y}\right]$$
$$= [\mathbf{u_i} \cdot \nabla \mathbf{P_i}] + \eta_{ix}\left(\frac{\partial v_{ij\|}}{\partial x}\right)^2 + \eta_{iy}\left(\frac{\partial v_{i\|}}{\partial y}\right)^2 + K_{qj}(T_e - T_i) + \frac{1}{2}m_i v_{i\|}^2 n_i \nu_{iz} + S_{Ej}$$

-------(6)

with $\eta_{x,y}$, $\kappa_{x,y}$, and $C_{x,y}$ as described above. $\nu_{iz}$ is the collision frequency between i and z, where the subscripts i and z refer to the charge of the main and impurity species, respectively. For the present case, pure hydrogen plasma z=1. The terms in square brackets are ion heat fluxes in respective directions ($q_{x,i}$ and $q_{y,i}$) and analogous to the electron heat fluxes. $S_{Ej}$ are other volumetric ion energy sinks and sources, such as neutral beam injection (NBI) and radio-frequency (RF) heating.

The equations for neutral is as follows: The continuity equation for hydrogen atom is

$$\frac{\partial}{\partial t}n_g + \frac{\partial}{\partial x}(n_g v_{gx}) + \frac{\partial}{\partial y}(n_g v_{gy}) = -<\sigma_i v_{te}> n_e n_g + <\sigma_r v_{te}> n_e n_i,$$

------(7)

where g stands for neutral hydrogen. The parallel momentum of neutral hydrogen is:

$$\frac{\partial}{\partial t}(m_g n_g v_{g\|}) + \frac{\partial}{\partial x}\left(m_g n_g v_{g\|} u_{gx} - \eta_{gx}\frac{\partial v_{g\|}}{\partial x}\right) + \frac{\partial}{\partial y}\left(m_g n_g v_{g\|} u_{gy} - \eta_{gy}\frac{\partial v_{g\|}}{\partial y}\right)$$
$$= \frac{B_x}{B}\left(-\frac{\partial P_g}{\partial x}\right) - m_i n_g \nu_{cx}(v_{i\|} - v_{g\|}) - m_g n_g \nu_i v_{g\|} + m_i n_i \nu_r v_{i\|}$$

------(8)

The details of the parallel and perpendicular component of neutral hydrogen velocities are given in Ref [15].

In the UEDGE code, the current continuity equation $\nabla \cdot J(\Phi)=0$ is used to calculate the potential [15]. In the present work potential equation is not considered so the details of the potential equation are not mentioned. In the present simulation, the drift terms are switched off. As a result, the poloidal ion velocity has only the first term on the R.H.S of equtaion (2) and radial ion velocity has first and second terms of equation (3), respectively. For the present case, $S_{Ee}$ and $S_{Ej}$ which are the last term of the R.H.S. of equations (5) and (6), respectively, are also omitted.

**3.2 Inclusion of limiter geometry if ADITYA-U in the UEDGE code**

The UEDGE code is primarily developed for simulating the behavior of plasma particles in the edge region of fusion devices containing magnetic null points (Diverter-X points). The UEDGE code requires a discretized representation of the physical domain, which is known as a computational grid or computational mesh. The inbuilt mesh generator routine in the UEDGE code using MHD equilibria, generates the computational mesh for a divertor geometry having X-points. Insertion of additional limiter structure up to the LCFS in the divertor configuration is also possible by this inbuilt mesh generator routine. However, mesh generation for a circular plasma without the null points is not possible with the inbuilt routine. Therefore, a separate mesh-generator routine is developed on MATLAB/PYTHON platform, based on plasma equilibrium flux surfaces obtained from equilibrium code IPREQ [26], to generate the mesh structures for the ADITYA-U limiter configuration. The mesh generator routine creates a computational mesh with the help of closed and open magnetic flux surfaces at the core-edge and SoL regions, respectively. The mesh structure is comprised of a collection of quadrilaterals (cells) associated with four vertices and one centre having R and Z coordinates. It is worth mentioning that these coordinates must be linked with the adjacent cells so that the core-edge region connects the SoL region. Figure 1 illustrates the formation of cells on the magnetic flux surfaces. Four sides A, B, C, and D define a quadrilateral cell of a mesh, where AB is the top-face, BD is the right-face, the CD is the bottom-face, and CA is the left-face of the cell. The poloidal sides (AC and BD) of the quadrilateral indicate two different poloidal surfaces of flux $\Psi(R, Z) = \Psi_1$ and $\Psi(R, Z) = \Psi_2$. The poloidal sides are constructed in such a way that one set of mesh surfaces becomes nearly orthogonal to the magnetic flux surfaces. During the translation of the points C and D on the poloidal flux surface, the overlapping with other neighboring points is avoided. The size of these cells is chosen to fulfill the following criteria. The number of cells for the core region needs to be the same as that of the SOL region to ensure continuity of the solution across the separatrix. It is worth mentioning that, the connectivity between the core-edge and SOL regions is straightforward because cells form continuous mesh lines that start from one surface of the limiter and end on the other surface of the limiter.

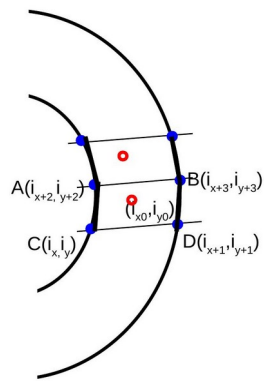

**Figure 1.** Generation of mesh structures using equilibrium flux surfaces. A, B, C and D are four corner points, whereas ($i_{x0}$,$i_{y0}$) is the centre.

Ninety-three cells are taken along each magnetic flux surface, whereas sixteen cells are considered radially across the magnetic flux surfaces. It was well explored in the past that various parameters associated with plasma and neutral fluid change sharply near the solid (limiter/divertor) surfaces [25,40] which is linked to strong recycling. Therefore, to simulate these parameters precisely using UEDGE, the size of the poloidal cell is expected to be much smaller compared to the mean-free paths of the atomic reaction processes, such as ionization, charge-exchange, etc. For typical discharges of ADITYA-U tokamak, the mean free path ($\lambda$) for ionization becomes $\lambda=(v_n/(ne*s_1))$ ~ 5 cm (where $v_n$ is the thermal velocity of neutral=1.5 x $10^3$ cm/s for a neutral temperature of ($t_n$)=0.025 eV and $s_1$ is the ionization rate coefficient) [41]. In addition, it has also been observed that there exists an abrupt change in the energy and particle loss along the parallel direction at the limiter tip due to a sudden change in the configuration of magnetic flux lines. It essentially creates strong radial gradients in the parameters associated with the plasma and the neutral fluid. Hence, the radial cell size is made sufficiently small to account for these sharp gradients. So, in the non-uniform mesh structure, the minimum poloidal cell arm is set to be ~ 5 mm, whereas the radial arm is made to be ~ 1 mm near the limiter surfaces. The wedge angle between the limiter surfaces is kept at ~ 80°. The mesh structure is then successfully integrated into the UEDGE code to estimate the edge plasma parameters. Figure 2 shows the two-dimensional non-uniform mesh structure for ADITYA-U tokamak plasma, whereas Figure 3 shows the zoomed view of the mesh near the limiter surface.

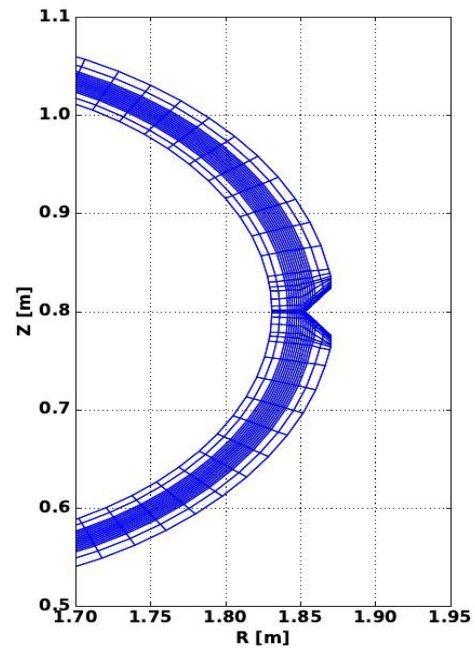

**Figure 2.** Computational mesh for ADITYA-U tokamak with limiter configuration.

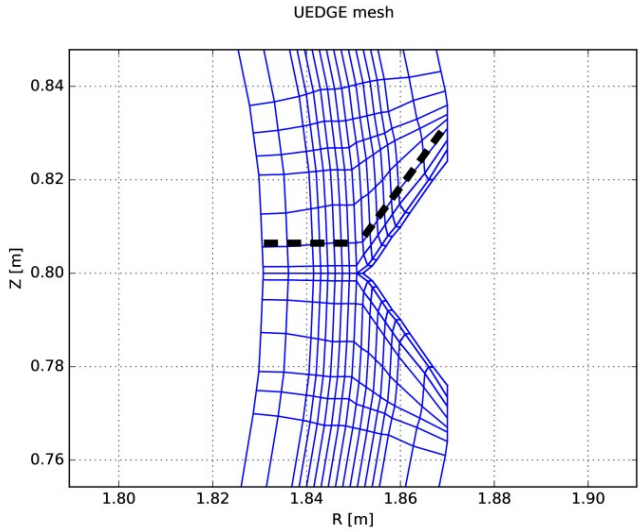

**Figure 3.** Zoomed view of mesh near the limiter structure. All the simulated parameters (as shown in figures 5-9) are obtained along the black dashed line which is 5 mm vertically above from the limiter tip.

### 3.3 Boundary conditions

At the innermost core-boundary, 2.0 cm inside the separatrix, the ion density $n_i=3.5\times10^{18}$ m$^{-3}$ is fixed. Electron ($T_e$) and ion ($T_i$) temperatures are also set to constant values ($T_e=T_i=$ 40 eV) and the parallel velocity is set to zero radial derivative ($du_\parallel/dy=0$) at this location. The plasma flows to a perfectly absorbing wall which is recycled as neutrals. At the limiter plate boundaries, $u_\parallel=c_s$. The electron and ion energy transmission factors are 5.0 and 3.5, respectively. Sputtering effects are not included. The limiting condition for the parallel transport of heat flux (used in equation (5) ) can be given by:

$$\chi = \frac{\chi_s}{[1+|q_s/q_f|^\beta]^{1/\beta}}$$

where $\chi_s$ is the classical diffusion transport coefficient and $\beta=1$. qs is the classical Spitzer heat flux. The second heat flux, $q_f$ is the free-streaming flux which is defined by $q_f = 0.21 *n_e *v_{te} *t_e$ for electron thermal transport, where 0.21 is a constant parameter that is taken by matching with kinetic modeling of transport [15]. The ion heat flux is limited in the same way as the electron heat flux. The ion parallel stress is limited to $fac*n_i*t_i$, where fac is the ion parallel viscosity coefficient, which is set to 1.

The simulations are run time-dependently with increasing time-step using a Krylov-Newton solver to achieve steady-state convergence. The initial time step which is used for the present simulation is $10^{-9}$ sec and the simulations evolved time-dependently at increasing time steps until a steady state is achieved with the norm of the residual less than $10^{-8}$. The other statistical data for a typical run is as follows: The total simulation time is 1 sec, the total number of time steps is 100, the number of linear iterations is 14 and the non-linear iteration is 1. The number of preconditioners solves is 15. Furthermore, the norm of the residual is 33 at the beginning and it is reduced to ~$10^{-9}$ at the end of the run.

## 4. Results and Discussions

The edge-region of hydrogen circular plasmas of ADITYA-U tokamak with limiter configuration is simulated with the help of the modified UEDGE code to model the experimentally measured electron density profile of discharge # 34166. The typical discharge parameters are shown in Figure 4.

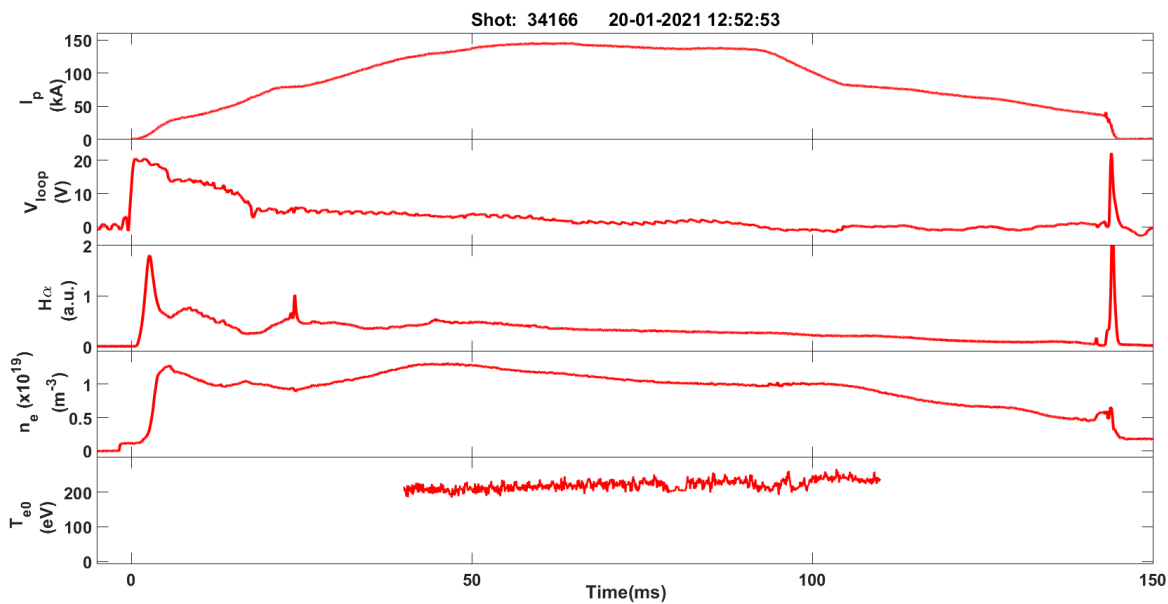

**Figure 4.** Typical discharge of ADITYA tokamak (a) plasma current (b) loop voltage (c) $H_\alpha$ signal (d) electron density (e) electron temperature.

It has maximum plasma current, $I_p$ of 120 kA and plasma duration of 145 ms. The maximum central chord averaged electron density, $n_e$ of $1.3 \times 10^{19}$ m$^{-3}$ and the core plasma electron temperature, $T_e$ of ~ 270 eV. Applying the boundary conditions as discussed in subsection 3.3, the code is primarily set-up using the following inputs for modelling the measured electron density profile. The density and temperature value at the core-edge is taken to be ~ $3.5 \times 10^{18}$ m$^{-3}$ and ~ 40 eV, respectively. The radial electron and ion heat diffusivities are as $\chi_e = \chi_i = 0.2$ m$^2$/s. The background neutral particle source is considered and its value is ~ $10^{15}$ m$^{-3}$. Here, the simulated radial profiles of electron density is obtained at a poloidal location above the limiter-tip. The limiter-tip (coordinates: R = 1.846 m, Z = 0.8 m as shown in Figure 3) demarcates the core-edge and SOL regions. Initially, the simulation is done by considering only radially constant perpendicular diffusion coefficient, D and without invoking the inward convective velocity, $v_{conv.}$ in the code. However, it is found that the simulated $n_e$ profile could not be properly matched with experimental data by using only perpendicular diffusive transport. Figure 5 shows the simulated radial profiles of electron density over a range of diffusion coefficients of 0.2, 0.5 and 1.0 m$^2$/s in absence of inward convective velocity. In this figure, the horizontal axis represents the x-projection of the line as shown by the black dashed line in Figure 3 and situated 0.5 cm vertically above the limiter tip. The '0' on the x-axis shows the position of the limiter tip. It is found that for only diffusive transport the mis-match between the simulated and measured profile increases particularly in the SOL region (r > $r_{limiter-tip}$) as shown by the black solid, blue dashed and red dash-dotted lines in Figure 5. It is worth mentioning that, the linear variation of D values (linearly increasing or decreasing within the range 0.2 to 0.6 m$^2$/s) with radial locations predict similar electron density profile as with the constant D =0.2 m$^2$/s. Also, for smaller value of D i.e. for 0.2 m$^2$/s, the electron density shows a bulge around -0.5 cm, inside the edge plasma and it also increases towards the outer wall region. When the value of D was increased from 0.2 to 1.0 m$^2$/s, the density profile was flatten as shown by blue dashed and red dash-dotted lines. This led to the conclusion that non diffusion transport in terms of convection velocity is needed to be included in the simulation to match the experimental observations with the simulated values.

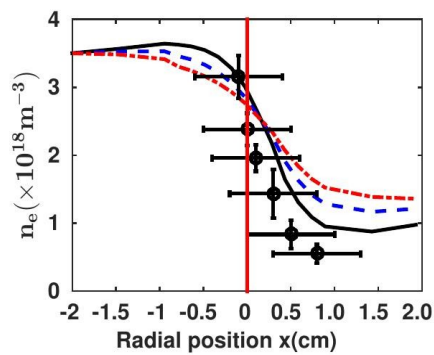

**Figure 5.** Simulated radial profiles of electron density for various diffusion coefficients 0.2, 0.5 and 1.0 m$^2$/s and without v$_{conv.}$ along the x projection of the line (As shown by the black dashed line in Figure 3). The open circles with error bars are the experimental data for typical discharge of ADITYA-U tokamak. The black solid, blue dashed and red dash-dotted curves represent the simulated radial profile of electron density for D =0.2 m$^2$ /s , 0.5 m$^2$ /s and 1.0 m$^2$ /s , respectively.

Then, non-diffusive transport, the anomalous inward convective velocity (v$_{conv.}$), is incorporated into the code to model the experimental n$_e$ profile. The values inward v$_{conv.}$ are varied through the iteration for obtaining the best-matched condition between simulated and experimental n$_e$ profile. Not only that, simulation was done with various combination D and v$_{conv.}$ values for obtaining the best match between experimental and simulated n$_e$ profiles. The variation of radial electron density profile for various combinations of inward convective velocities at a fixed value of D of 2.0 m$^2$/s are simulated and shown in Figures 6 and 7. Figure 6(a) shows the radial profiles of electron density for four different step functions of inward v$_{conv.}$ with different inward convective velocities: case-I is a constant 1.5 m/s throughout the whole region (black solid line), case-II and Case-III is with v$_{conv.}$ = 1.5 m/s (blue-dashed line) and v$_{conv.}$ = 5.0 m/s (black dashed-dotted line) at r < r$_{limiter-tip}$ and almost zero value at SOL region and case-IV is having 1.0 m/s at r < r$_{limiter-tip}$ region and higher value of 5.0 m/s at SOL region (red solid line with tringles). Among them, Case-II and case-III predict the similar density profiles (blue-dashed line and black dashed-dotted lines) as shown in Figure 6(a). Here, the simulated n$_e$ profile shows a tendency of bulging up around 0.5 cm inside the edge region as observed for only constant D case. Furthermore, for case-IV, which is having the lower value of 1.0 m/s at r < r$_{limiter-tip}$ region and higher value of 5.0 m/s at SOL region, the simulated n$_e$ profiles coincide with the profile with constant value of v$_{conv.}$ =1.5 m/s as shown in black solid line in Figure 6(a). This step function having higher v$_{conv.}$ at SOL region was varied for several values starting from 1.5 (i.e basically constant v$_{conv.}$ profile) to 15 m/s. It is found that the matching between simulated and experimental data are almost similar to the case of constant v$_{conv.}$ and that reveal that variation of the magnitude of the constant inward convection velocity inside the SOL region does not play important role in the modelling of n$_e$ profile in ADITYA-U tokamak. But it is strongly affected by the step functions of opposite nature having lower inward v$_{conv.}$ in SOL as compared to their values of inside the LCFS as shown by blue-dashed line and black dashed-dotted lines.

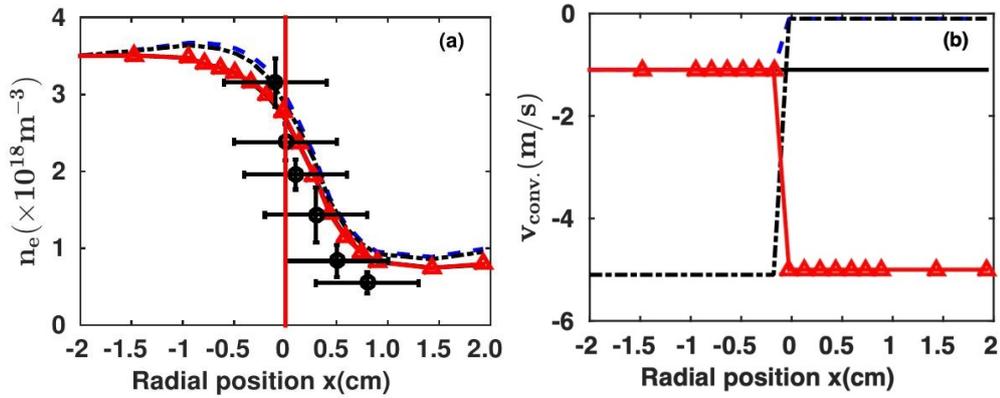

**Figure 6** Simulated radial profiles of (a) electron density for diffusion coefficients 0.2 m²/s with (b) various inward $v_{conv.}$ along the x projection of the line (As shown by the black dashed line in Figure 3). The open circles with error bars are the experimental data for typical discharge of ADITYA-U tokamak.

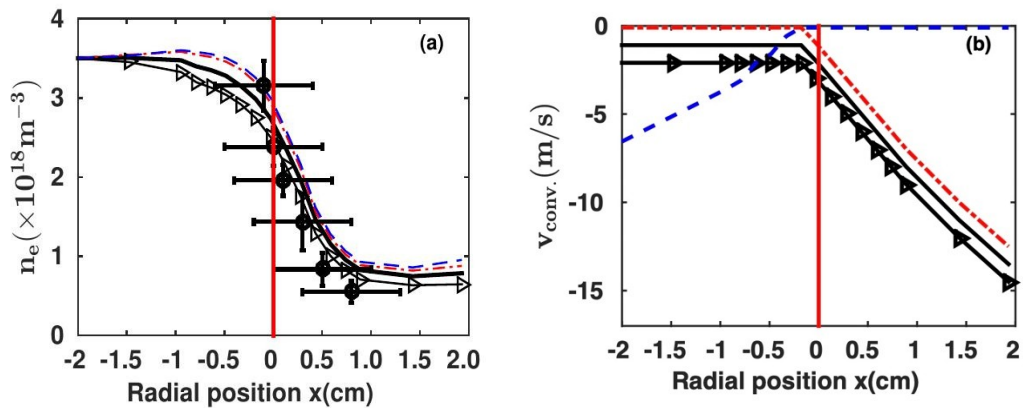

**Figure 7** Similar to Figure 6 (a) electron density for diffusion coefficients 0.2 m²/s with (b) various inward $v_{conv.}$

Not only that, additional inward convection velocity profiles, such as straight line and exponential natures growing towards the outer wall, were also considered for matching. Figure 7(a) illustrates the outcomes of the convection velocity's radially increasing nature in the SOL region, and Figure 7(b) displays the corresponding inward convective velocities. Among them, the black solid line in

Figure 7(b) represents for the constant convective velocity till $r < r_{limiter-tip}$ and after that it increases monotonically to a value 14 m/s in the SOL region. The simulated density profile for this case remains almost same as of the $v_{conv.}=1.5$ m/s as delineated in Figure 6(a) by black solid line. The black solid line with triangles and red dashed line represent a couple of convective velocities, which lie below and above the black solid line (see Figure 7(b)), respectively, for which the electron densities are simulated. It is observed from Figure 7(a) that in the case of black solid line with triangles profile the extent of mis-match between the simulated and measured electron density profiles increases in the core-edge region ($r < r_{limiter-tip}$) as compared to SOL region. This further suggest that the values of $v_{conv.}$ inside the core-edge region play important role towards matching with experimental $n_e$ profile. Furthermore, the more decreased values of $v_{conv.}$ at $r < r_{limiter-tip}$ is also taken into account to simulate the electron density and it is noticed that the density profile (blue-dashed line) remains similar in nature as of the red-dashed line (see Figure 7(a)). Both the cases of $v_{conv.}$ profile represented by red dashed line and the profile with gradually increasing inward $v_{conv.}$ within $r < r_{limiter-tip}$ and almost by keeping zero value within SOL (blue dashed line) shows the larger mis-match between simulated and experimental data. From this study it can be inferred that the nature of the edge electron density does not depend on the inward convective velocity if it is taken to be a constant of 1.5 m/s at core edge region ($r < r_{limiter-tip}$) and for a value of $v_{conv.} >=1.5$ at SOL region ($r > r_{limiter-tip}$), but it strongly depend on the the values of $v_{conv.}$ inside the core-edge region.

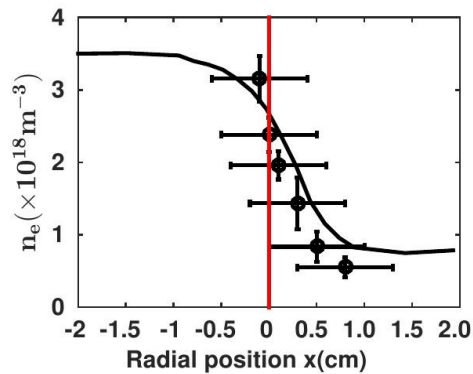

**Figure 8** Simulated radial profiles of electron density for diffusion coefficients 0.2 m²/s with inward $v_{conv.}=1.5$ m/s along the x projection of the line (As shown by the black dashed line in Figure 3). The open circles with error bars are the experimental data for typical discharge of ADITYA-U tokamak.

Finally, the best-matched simulated and experimental radial profiles of $n_e$ are shown in Figure 8. It is found that a combination of perpendicular particle diffusion coefficient ~ 0.2 m²/sec and a radially constant inward convective velocity of 1.5 m/s are required for obtaining the best-matched condition as shown with the black solid line in Figure 8. As reported in Ref. [36] that ware pinch is responsible for the inward plasma movement. Following the above mentioned reference, the Ware pinch velocity, i.e. $E_\Phi/B_\theta$ is estimated from the measured toroidal electric field, $E_\Phi \sim 0.38$ V m$^{-1}$ (loop voltage $\sim 2$ V as shown in Figure 4) and $B_\theta \sim 0.1$ T, comes out to be $\sim 3$ m s$^{-1}$, which is approximately same as of the assumed value of $|v_{conv.}| = 1.5$ m s$^{-1}$. Although not in many tokamaks, the existence of strong convection in the far edge and SOL regions of the plasma is also observed in DII-D tokamak [21]. It has been found that the edge plasma parameters are matched well with the experimental measurements in SOL and divertor region after considering the maximum convective velocity in the order of 50 to 100 m/s along with the diffusion coefficient in the range of 0.125-0.3 m$^2$/s depends upon various plasma conditions of DIII-D. The cross-field transport coefficients along with the spatially varying convective velocity in UEDGE code are considered to predict the midplane density profiles, target plate heat flux profiles, and inner/outer divertor power sharing for SPARC [42].

The D value $\sim 0.2$ m$^2$/sec obtained from matching the simulated and measured density profile agrees well with those estimated through the analytical formula $D = (\lambda_{SoL}^2 \, c_s)/L$ [1]. Taking, L ($=\pi Rq$) $\sim 10$ m, $c_s = 3.54 \times 10^4$ m/s and $\lambda_{SoL} \sim 8$ mm calculated using $n = n_0 \exp(-x/\lambda_{SoL})$ from the measured radial density profile yielded a value of 0.2 m$^2$/s. Furthermore, to understand the nature of the edge particle transport, the diffusion coefficient is calculated using neo-classical transport. The neo-classical transport of particles at the plasma edge of ADITYA-U falls in the Pfirsch-Schluter regime as it satisfies the condition; Pfirsch-Schluter frequency, $\nu_{PS}$ ($=v_{Ti}/Rq$) $\sim 1 \times 10^4$ s$^{-1}$ < $\nu_{coll} \sim 2. \times 10^5$ s$^{-1}$, where $\nu_{coll}$ is the ion-ion collision frequency [43] and $v_{Ti} \sim 3.0 \times 10^4$ m/s is the ion thermal velocity. The value of $D_{PS}$ comes out to be 0.02 m$^2$/s, which is one order lower than that of the simulated value. The Bohm diffusion [30] ($D_B = 1/16 \, (T_e/B)$) m$^2$/s, where $T_e$ is in eV and B is in Tesla) the coefficient comes out to be $\sim 1$ m$^2$/s for $T_e = 15.0$ eV and B = 1 T, which is almost five times higher than the simulated value. Although the simulated particle perpendicular diffusion coefficient is observed to be higher than the neo-classical value, they are found to be significantly less than those obtained in the case of fluctuation induced transport. The diffusion coefficient due to ITG mode is given by [31]

$$D_{ITG} \sim \frac{c_s \rho_s^2}{(L_p R)^{1/2}} \left(\frac{q^4 R}{L_p}\right)^{(1/4)}.$$

Where $c_s=(T_i+T_e)/m_i$ and $\rho_s=c_s/\Omega_{ci}$ are the velocity and length traversed by the ion sound waves, respectively, $\Omega_{ci}$ is the ion cyclotron frequency. Taking electron temperature ($T_e$) ~ 15 eV, ion temperature ($T_i$) ~ 5 eV and electron density ($n_e$) ~ 3 x $10^{18}$ m$^{-3}$, $q_{edge}$ ~ 4, the $D_{ITG}$ ~ 1.5 m$^2$/s. The estimated $D_{ITG}$ is approximately seven times higher than the simulated value. Furthermore, the diffusion coefficient is estimated from resistive ballooning mode using the expression

$$D_{RB} = C_{RB}(2\pi q)^2 \nu_{ez}\rho_e^2 \left(\frac{R}{L_p}\right).$$

Where the value of $C_{RB}$ is chosen between 10 and 30 [32] and other symbols have their usual meaning. Taking the above-mentioned edge plasma parameters of ADITYA-U plasma, $D_{RB}$ comes out to be ~ 320 m$^2$/s, which is significantly higher than the simulated value. Hence fluctuation induced transport does not seem to be influencing the particle diffusivity in the edge region of the ADITYA-U plasma.

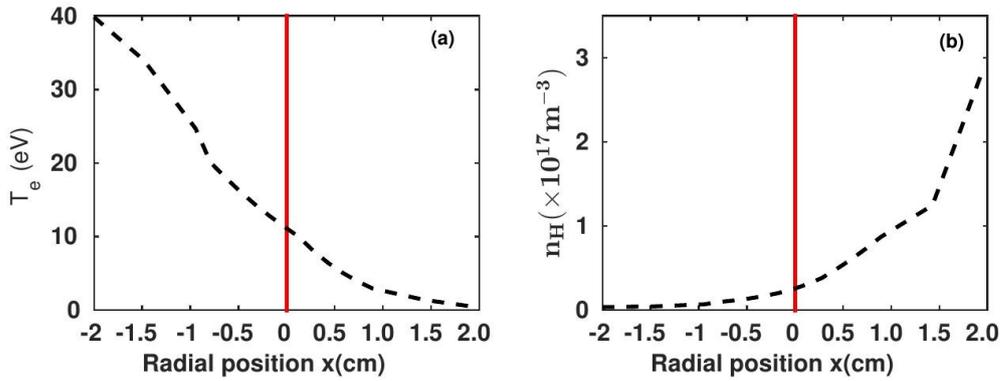

**Figure 9.** Variation of (a) electron temperature and (b) neutral hydrogen density along the x-projection of the line (As shown by the black dashed line in Figure 3) situated 5 mm vertically above from the limiter tip. The '0' indicates the position of the limiter tip. The open circles with error bars are the experimental data for typical discharge of ADITYA-U tokamak.

Figures 9 (a) and (b) show the electron temperature, $T_e$ and neutral hydrogen density, $n_H$ which are the other outcomes of the UEDGE code. It is noticed that, near the limiter tip location, $T_e$ is around 12 eV and inside the rlim=0 the temperature is increased upto 40 eV (See, Figure 7(a)) which is the value of fixed core-edge temperature as discussed in first paragraph in section "Results and Discussions". Since the experimentally measured temperature is available for only one radial location which is near the limiter tip location and it is of ~ 12 eV. It is found that the simulated temperature value at this location is matches well with the experimental measurements [36]. Figure 9(b) shows the simulated neutral hydrogen density ($n_H$) from UEDGE code. The $n_H$ maximizes in the SOL and decreases rapidly towards the limiter tip to almost one order less than its maximum value within ~ 2 cm from the outer wall. Furthermore, the estimated radial profile of heat flux due to conduction and convection are also obtained for the matched condition of simulated and measured density profiles as shown in Figure 5. Figure 10 shows the conductive and convective heat fluxes for electrons and ions. It has been found that the maximum radial conductive heat flux of electron which is calculated using $n_e \chi_e \nabla T_e$ (As defined in electron energy equation (equation (5)) is ~$2.1 \times 10^{21}$ eV m$^{-2}$ s$^{-1}$ at ~ 0.94 cm inside the limiter location. The electron temperature profile is as shown in Figure 7(a). Whereas, the maximum radial convective heat flux for the electron (=2.5 $T_e \Gamma_{ne}$=2.5 $T_e$ (-D d$n_e$/dr+$v_{conv.}$ $n_e$)) ~ $1.3 \times 10^{21}$ eV m$^{-2}$ s$^{-1}$ is observed near the limiter tip. The radial conductive heat transport is dominated over the radial convective heat transport from core-edge (r~-2 cm) to ~ - 0.5 cm inside the limiter location. The radial conductive heat flux is more than the convective heat flux because, radial convective velocity term is dominated by density gradient while the conductive heat flux term contains the temperature gradient and in the region -2.0 cm < r < -0.5 cm (see, Figure 10), the density scale length is higher than the temperature scale length. Similarly, the radial profiles of conductive and convective heat fluxes for ions (As defined in ion energy equation (equation (6))) are obtained from the simulations. The simulated ion temperature profile from UEDGE code is plotted in Figure 11 which is used in the calculations of radial profiles of conductive and convective heat fluxes for ions. It shows that the conductive part for ion is higher than the electron heat flux in the region -2.0 cm < r < -0.5 cm, whereas the convective part is always lower than the electron heat flux. The total radial electron heat flux ~ $0.34 \times 10^{22}$ eV m$^{-2}$ s$^{-1}$ near the edge-SOL transition is obtained from the UEDGE simulation. This value is close to those obtained from the BOUT++ simulation for ADITYA-U edge plasma parameters within the limitation of boundary values [44].

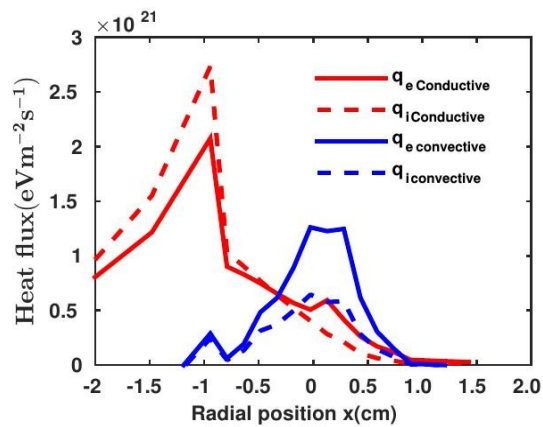

**Figure 10.** Simulated radial profiles of heat-flux components of equations (5) and (6) for electrons (Solid curve) and ions (dashed curve), respectively, along with the x projection of the line (As shown by the black dashed line in Figure 3). The red curve represents radial heat conductivity and the blue curve represents radial heat flux due to convection using the simulated temperature profile of electron and ion as shown in Figures 9 (a) and 11, respectively).

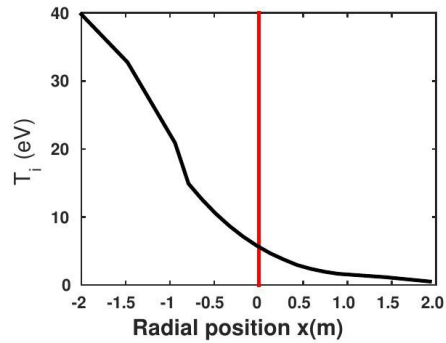

**Figure 11.** Simulated radial profile of temperature for ions along with the x projection of the line (As shown by the black dashed line in Figure 3).

Figure 12 and Figure 13 depict the zoomed views of simulated 2D profiles of edge electron density and electron temperature near the limiter region, respectively. The simulation showed that the electron density falls almost four times from its value at the core-edge region to the SoL region. Further, as shown in Figure 12 that in the region of open field lines, i.e., $r > r_{limiter-tip}$, where the field lines impinge on the limiter surfaces from opposite directions, the ions stream onto the limiter surface from both sides, but with opposite poloidal velocities on the upper and lower sides of the limiter surface [25]. Whereas, in the region where the magnetic field lines are closed i.e., $r < r_{limiter-tip}$, the plasma flows poloidally in only one direction. Figure 13 shows the 2D variation of electron temperature from core edge to SOL region, with the temperature at the core-edge boundary ~ 40 eV falling to 0.2 eV at the wall.

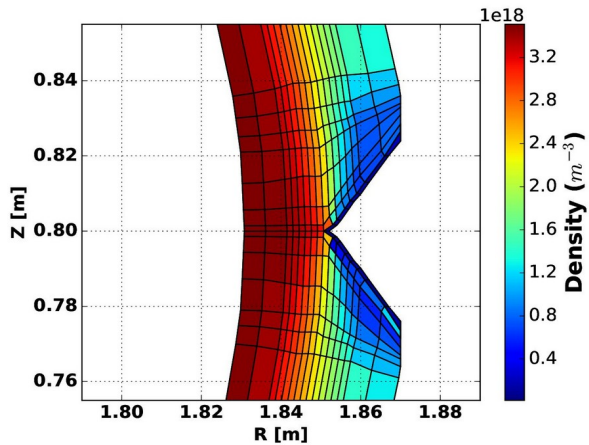

**Figure 12.** Simulated 2D profile of electron density. The coordinates of the limiter tip are: R=1.846 m, Z=0.8 m.

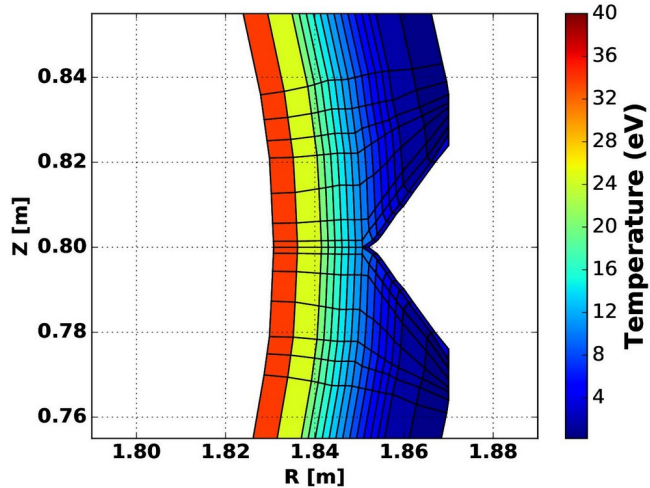

**Figure 13.** Simulated 2D profile of electron temperature. The coordinates of the limiter tip are: R=1.846 m, Z=0.8 m.

## 5. Conclusion

The edge region of circular limiter plasmas of ADITYA-U is simulated using the fluid transport code, UEDGE to model the ADITYA-U experimental measurements. The UEDGE code, which has primarily been developed and commonly used in tokamaks with divertor geometry, is modified to be used in limiter configuration. A routine has been developed to generate the computational mesh defining the limiter geometry of ADITYA-U using equilibrium magnetic flux surfaces obtained from the IPREQ code. The mesh-generator routine is successfully integrated into the UEDGE code and the edge plasma parameters such as electron density and electron temperature are simulated. The Krylov-Newton solver is used for the code to run time-dependently until a steady-state solution. It is found that using only perpendicular diffusive transport, the simulated radial profile of $n_e$ could not be properly matched with experimental data measured by Langmuir probe. However, the simulated radial profile of $n_e$ matches well with the measurements, when a constant perpendicular diffusion coefficient, D ~ 0.2 $m^2$/s and a radially constant convective velocity have been incorporated in the simulation. The value of particle diffusion coefficient (D) ~ 0.2 $m^2$/s is found to be larger than the neoclassical value but smaller than the estimated Bohm diffusion coefficient and significantly smaller than those estimated with fluctuation induced transport. Hence, fluctuation induced particle transport is less likely in the edge region of ADITYA-U tokamak. In addition, the conductive and convective radial heat fluxes show a maximum value of ~ $2.1 \times 10^{21}$ eV m$^{-2}$ s$^{-1}$ and ~ $1.3 \times 10^{21}$ eV m$^{-2}$ s$^{-1}$, respectively, for electrons. The convective heat transfer mechanisms contribute maximum to the total heat transfer near the limiter-tip location. The successful setting up of the UEDGE code to obtain edge radial profiles of electron density and temperature is a step forward for coupling it with the already established DEGAS2 code [27, 28, 29] to simulate the neutral dynamics in ADITYA-U tokamak.

6. Acknowledgements

With profound grief, the authors acknowledge the contribution of late Dr. R. Srinivasan (we lost him in the COVID19 pandemic) in this work, who has developed the IPREQ code in IPR. One of the authors (Dr. Ritu Dey) is thankful to Mr. Shirsh Raj for helpful Discussions.